\documentclass[%
 superscriptaddress,
 preprint,
 amsmath,amssymb,
 floats,
]{revtex4-1}

\usepackage{graphicx}
\usepackage{dcolumn}
\usepackage{bm}
\usepackage{bm, color}

\begin{document}


\title{Strong spin-Hall and Nernst effects in a $p$-band semimetal}

\author{Yang Zhang}
\thanks{These two authors contributed equally}
\affiliation{Max Planck Institute for Chemical Physics of Solids, 01187 Dresden, Germany}
\affiliation{Leibniz Institute for Solid State and Materials Research, 01069 Dresden, Germany}

\author{Qiunan Xu}
\thanks{These two authors contributed equally}
\affiliation{Max Planck Institute for Chemical Physics of Solids, 01187 Dresden, Germany}

\author{Klaus Koepernik}
\affiliation{Leibniz Institute for Solid State and Materials Research, 01069 Dresden, Germany}

\author{Johannes Gooth}
\affiliation{Max Planck Institute for Chemical Physics of Solids, 01187 Dresden, Germany}

\author{Jeroen van den Brink}
\affiliation{Leibniz Institute for Solid State and Materials Research, 01069 Dresden, Germany}

\author{Claudia Felser}
\affiliation{Max Planck Institute for Chemical Physics of Solids, 01187 Dresden, Germany}

\author{Yan Sun}
\email{ysun@cpfs.mpg.de}
\affiliation{Max Planck Institute for Chemical Physics of Solids, 01187 Dresden, Germany}

\begin{abstract}
Since spin currents can be generated, detected, and manipulated via the 
spin Hall effect (SHE), the design of strong SHE materials has become a 
focus in the field of spintronics. 
Because of the recent experimental progress also the spin Nernst effect 
(SNE), the thermoelectrical counterpart of the SHE, has attracted much interest.
Empirically strong SHEs and SNEs are associated with $d$-band compounds, such 
as transition metals and their alloys -- the largest spin Hall conductivity 
(SHC) in a $p$-band material is 
$\sim 450$ $\left(\hbar/e\right)\left(\Omega\cdot cm\right)^{-1}$ for a 
Bi-Sb alloy, which is only about a fifth of platinum. 
This raises the question whether either the SHE and SNE are 
naturally suppressed in $p$-bands compounds, or favourable $p$-band 
systems were just not identified yet.
Here we consider the $p$-band semimetal InBi, and predict it has a 
record SHC $\sigma_{xy}^{z}\approx 1100 \ \left(\hbar/e\right)\left(\Omega\cdot cm\right)^{-1}$ 
which is due to the presence of nodal-lines in its band structure. 
Also the spin-Nernst conductivity $\alpha_{zx}^y\approx 1.2  \ (\hbar/e)(A/m\cdot K)$ 
is very large, but our analysis shows its origin is different as the maximum appears 
in a different tensor element.
This insight gained on InBi provides guiding principles to obtain a strong SHE 
and SNE in $p$-band materials and establishes a more comprehensive understanding 
of the relationship between the SHE and SNE.
\end{abstract}

\maketitle

The joint utilisation of the spin and charge degree of freedom in solids
is the main target of spintronics, in which the spin current
generation, detection, and manipulation are the three crucial
objectives~\cite{Wolf2001, Zutic2004, Bader2010}. The spin Hall effect (SHE) provides an effective
technology to convert the charge current into a pure spin current, where
an applied electric field can generate a transverse spin current
with the spin polarization perpendicular to both the spin and charge 
currents~\cite{DYakonov1971, Hirsch1999, Kato2004,Sinova2015}.
Vice versa, the inverse SHE provides an effective method to manipulate
the spin current without a magnetic field~\cite{Hirsch1999,Valenzuela2006}. 
Due to this versatility the SHE has attracted extensive interest in recent years, and much effort has been devoted
to the theoretical understanding and engineering of strong SHE materials~\cite{Murakami2003,Sinova2004, Kato2004, Guo2005, Valenzuela2006, Tanaka2008,Hoffmann2013}.

In general the SHE has two distinct origins, an extrinsic contribution from the scattering
and an intrinsic one from the band structure.
The intrinsic part can be
formulated via the spin Berry curvature (SBC) similar to the anomalous Hall effect~\cite{Sinova2015,Nagaosa2010,Xiao2010}.
Apart from the SHE, a transverse spin current can be generated by a temperature
gradient instead of an electric field, which constitutes the spin Nernst effect
(SNE)~\cite{Meyer2017,sheng2017spin,kim2017observation}.
The SNE can also be formulated via the Kubo formula approach based on the electronic
band structure~\cite{Murakami2003,Sinova2004,Bernevig2005, Cheng2008spin,Bauer2012,
Xiao2010,tauber2012extrinsic,tauber2013spin,wimmer2013first}. Thus the electronic band structure plays a crucial
role in searching and designing strong SHE and SNE materials.

Though the precise reason is not clear to date, empirically materials with a strong SHE are primarily dominated
by $d$-orbital-related compounds~\cite{Tanaka2008, Hoffmann2013,Sinova2015, Kimura2007, Saitoh2006} such as 5$d$-transition
metals and alloys.
The largest spin Hall conductivity (SHC) found in $p$-band materials is in a Bi-Sb
alloy~\cite{Sahin2015,Fan2008} at only around 450 ($\left(\hbar/e\right)\left(\Omega\cdot cm\right)^{-1}$).
Very recently, it was found that the nodal line band structure could generate
strong SHCs because of the large local SBC~\cite{Sun2016, Sun2017}. Employing this guiding principle,
we theoretically predict a large intrinsic SHC of about $1100$ ($\left(\hbar/e\right)\left(\Omega\cdot cm\right)^{-1}$)
in the $p$-band semimetal InBi. Substitution of the electric field by a temperature gradient,
reveals a large spin Nernst conductivity (SNC) of about $1.2$ $(\hbar/e)(A/m\cdot K)$ at 300 K. This shows that the SNE and SHE can
be strongly enhanced by the topological band structure, and that the large SNC and SHC need not be intrinsically weak
in cheaper and therefore commercialy more attractive $p$-band materials. Though both the intrinsic SHE and SNE can be understood from the SBC, their largest value appears at different tensor elements, implying different origins.

\begin{figure}[htbp]
\begin{center}
\includegraphics[width=0.5\textwidth]{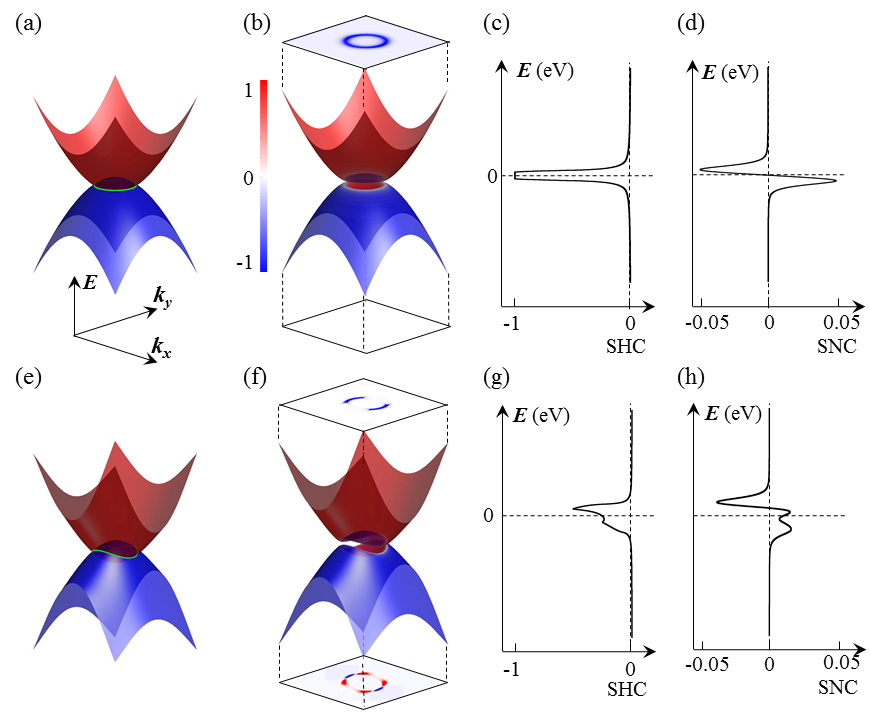}
\end{center}
\caption{
\textbf{Band structure, SHC, and SNC in the effective BHZ model.} (a) Nodal line energy dispersion of
the high symmetry effective BHZ model without the inclusion of SOC. (b) SOC breaks the nodal line with opening
a global band gap after SOC is taken into consideration. The distribution of SBC and SNBC at charge neutral point
are given above and below the energy dispersion. (c-d) Energy-dependent SHC and SNC, respectively.
(e-h) The same as (a-d) but for the symmetry-reduced model. The color bars are in arbitrary units.
}
\label{model}
\end{figure}

To understand the effect of the nodal line band structure on the SHE and especially the SNE,
we first consider two simple systems by effective model Hamiltonians. From the high-symmetry Benervig-Hughes-Zhang (BHZ) model for quantum spin Hall insulators~\cite{Bernevig_2006},
where $s_z$ is maintained as a good quantum
number (see the method section), the band structure is presented as a nodal line without
considering spin-orbit coupling (SOC). As soon as SOC is taken into consideration,
the linear band crossing is gapped and leads to a quantized spin Hall conductance
in the band gap~\cite{Sun2016}. Though the spin Hall conductance reaches the maximum value at the charge neutral point, the spin
Nernst conductance is zero there because of the electron and hole symmetry.
Comparing the SBC and spin Nernst Berry curvature (SNBC) distribution in reciprocal
space, it is found that the SNBC is zero at any $k$-point, while the SBC has a strong hot ring 
from the nodal-line-like band anti-crossing (see Fig. 1(b)). From Eq. (4), one can find that
the SNBC is actually a redistribution of the SBC because of the temperature effect, which should be
cancelled out by the electron-hole symmetry. To break the electron-hole symmetry, we have
reduced the symmetry of the effective model Hamiltonian, making the system non-insulating. Compared to the high-symmetry model, the nodal line in the symmetry-reduced model
has a dispersion in energy space, and the Fermi velocities are different along the $x$ and
$y$ directions. As there is no global gap, the spin
Hall conductance is not quantized anymore (Figs. 1(f--g)). Meanwhile, a finite spin
Nernst conductance appears at the charge neutral point. In Figs. 1(b,f), we also analysed the SBC
and SNBC distribution in the $k_x-k_y$ plane. Because of the dispersion of the nodal ring, the high
intensity of the SBC changes from a hot ring to two hotspots, resulting in a non-zero SNBC, as
presented in the lower panel of Fig. 1(f). By integrating the SNBC, one can obtain a non-zero spin
Nernst conductance at the charge neutral point. Therefore, the non-zero SNC must break the
balance of the SHC distribution in energy space.

\begin{figure*}[htbp]
\begin{center}
\includegraphics[width=0.5\textwidth]{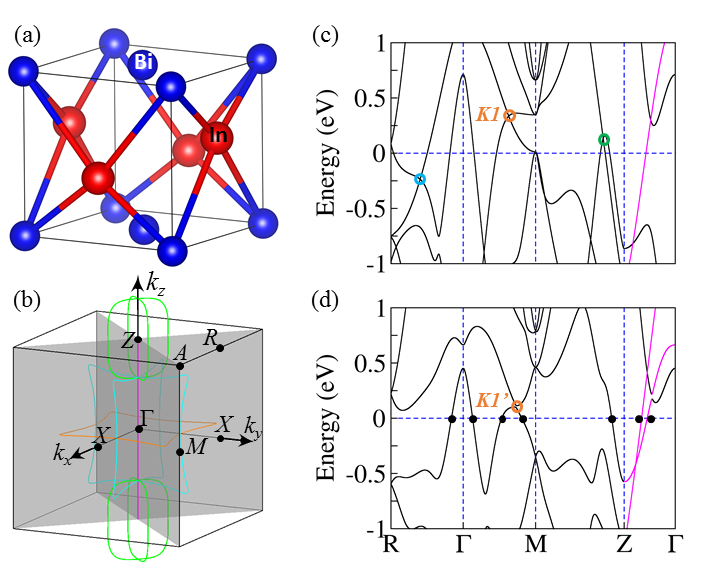}
\end{center}
\caption{
\textbf{Lattice and electronic band structure for InBi.}
(a) Tetragonal lattice structure in InBi with space group $P4/nmm$.
The lattice parameters are $a=b$=4.9846 $\AA$ and $c$=4.8116 $\AA$.
(b) Nodal line distribution in BZ. Here only the nodal lines
without the inclusion of SOC are shown, and they are gapped out by SOC.
(c-d) Energy dispersion without and with the inclusion of SOC. The nodal ring
linear band crossings in $k_z=0$, $k_{x,y}=0$, and $k_x \pm k_y=0$ planes,
are labeled by orange, cyan, and green circles, respectively, in (c).
The band anticrossing opened by SOC on the $\Gamma$-$M$ is also labeled by
an orange circle.
The nodal line on $\Gamma$-$Z$ is highlighted by magenta.
The transition point of band occupation are labeled by black dots in (d).
}
\label{lattic}
\end{figure*}

The specific material InBi has been reported to exhibit nonsymmorphic symmetry-protected
nodal lines at the edge of the Brillouin zone (BZ)~\cite{Sandy2017}. However, 
these kinds of nodal lines are just band degeneracies from band folding between 
different BZs. According to previous studies, this
kind of nodal line normally does not exhibit any topological charge or SBC and,
therefore, cannot generate the SHE and SNE~\cite{Sun2017}. In the current work, we find another type
of nodal line inside the BZ protected by mirror symmetry and rotation symmetry,
which generates a strong SHE and SNE. As shown in Fig. 2 (b), there are four classes of nodal lines, located
in the high-symmetry planes of $k_{x,y}$=0, $k_z$=0, and $k_{x}\pm k_{y}$=0, and high-symmetry line of $\Gamma-Z$, respectively. Without including the SOC, the inverted bands have
opposite mirror eigenvalues of 1 and -1, respectively, in the planes of $k_{x,y}$=0,
$k_z$=0, and $k_{x}\pm k_{y}$=0, leading to the double degeneracy (not considering
the spin degree of freedom)
of nodal line linear band crossings. Along the high-symmetry line of $\Gamma-Z$, the double band degeneracy
(highlighted by magenta in Fig. 2(c)) is due to the $c_4$ rotation symmetry with respect to the $z$-axis.
As long as SOC is taken into consideration, the spin rotation symmetry is broken and the linear band crossings
are also gapped out, as indicated in Figs. 2(c-d). This kind of band
anti-crossing
from the SOC is often accompanied by strong band entanglements and yields a large SBC.

\begin{figure}[htbp]
\begin{center}
\includegraphics[width=0.5\textwidth]{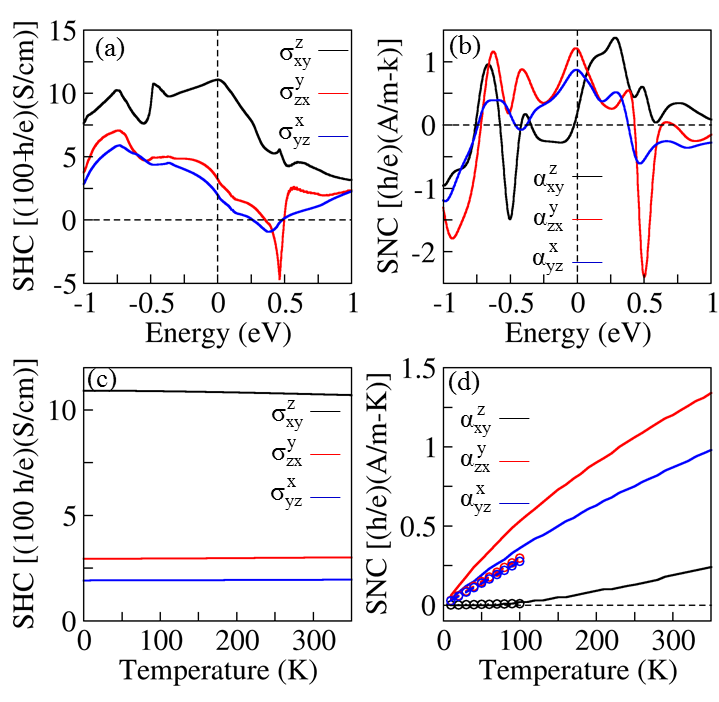}
\end{center}
\caption{
\textbf{SHC and SNC in InBi}.
(a, b) Energy dependent SHC and SNC for the three independent tensor
elements. (c,d) The evolution of SHC and SNC as the function of temperature.
The solid curves and empty circles in (d) are from equation (4) and
Mott relation, respectively.
}
\label{shc}
\end{figure}

The SHC($\sigma_{i,j}^{k}$) and SNC($\alpha_{i,j}^{k}$; $i,j,k=x,y,z$) are
$3\times3\times3$ tensors, representing the spin current
$\vec{J}_{si}^{k}$ generated by the electric field $\vec{E}$ via $\vec{Js}_{i}^{k}=\underset{j}{\sum}\sigma_{ij}^{k}\vec{E}_{j}$
and temperature gradient $\overrightarrow{\nabla T}$ by $\vec{Js}_{i}^{k}=\underset{j}{\sum}\alpha_{ij}^{k}\overrightarrow{\nabla T}_{j}$,
where $\vec{Js}_{i}^{k}$ flows along the $i$-direction with the spin-polarization along the $k-$direction, and $\vec{E}_j$ and
$\overrightarrow{\nabla T}_{j}$ are the $j-$component of the electric field $\vec{E}$ and temperature gradient $\overrightarrow{\nabla T}$,
respectively. Based on linear response theory, for the specific space group
$P4/nmm$, there are only three independent tensor elements
for both the SHC and SNC tensor~\cite{Kleiner1966, Seemann2015,zhang2017strong}:

\begin{equation}
\underline{X}^{x}=\left(\begin{array}{ccc}
0 & 0 & 0\\
0 & 0 & X_{yz}^{x}\\
0 & -X_{zx}^{y} & 0
\end{array}\right),\\
\underline{X}^{y}=\left(\begin{array}{ccc}
0 & 0 & -X_{yz}^{x}\\
0 & 0 & 0\\
X_{zx}^{y} & 0 & 0
\end{array}\right),\underline{X}^{z}=\left(\begin{array}{ccc}
0 & X_{xy}^{z} & 0\\
-X_{xy}^{z} & 0 & 0\\
0 & 0 & 0
\end{array}\right)
\end{equation}
where $X$=$\sigma$ and $\alpha$ represent the SHC and SNC, respectively. Therefore, there are only three independent
non-zero elements, following the relation of $X_{xy}^z$=-$X_{yx}^z$, $X_{zx}^y$=-$X_{zy}^x$, and $X_{yz}^x$=$X_{xz}^y$.

Our calculations are fully consistent with the symmetry analysis. From the energy-dependent SHC (Fig. 3(a)),
one can see that the largest tensor element appears at $\sigma_{xy}^z$, which can reach up to about $1100$
($\left(\hbar/e\right)\left(\Omega\cdot cm\right)^{-1}$). Thus far, this is the only SHC above 1000
($\left(\hbar/e\right)\left(\Omega\cdot cm\right)^{-1}$) in the reported
$p$-orbital compounds. This large value is robust in a large energy window from $E_0$--0.5 eV to the charge
neutral points in the hole-doped range. Therefore, hole doping is preferred from the SHE point of view.
Furthermore, we find that the SHC is robust with respect to temperature. The SHC only varies less than
$5\%$ from 0 K to room temperature, which is very similar to that in platinum~\cite{Meyer2017}

From the above model analysis, we already know that a large SHC does not imply a large SNC. To obtain a
large SNC, the balance of the SHC distribution in energy space must be broken. Based on this understanding, the robustness of the SHC in energy
space is not beneficial for the SNE, and a large SNC should not appear at $\alpha_{xy}^z$.
Indeed, the tensor element of
$\alpha_{xy}^z$ is very small, at about 0.2 $((\hbar/e)(A/m\cdot K))$ at 300 K. However, a quite large SNC
is achieved at $\alpha_{xy}^z$, which can reach up to about $1.2$ $((\hbar/e)(A/m\cdot K))$ at 300 K.
Because the SNE is very sensitive to temperature, we also analysed the evolution of the SNC as a
function of temperature. As presented in Fig. 3(c), $\alpha_{xy}^z$ is near-zero at low temperatures below
100 K, and the $\alpha_{zx}^y$ and $\alpha_{yz}^x$ are also small ($\sim0.5$ $((\hbar/e)(A/m\cdot K))$). All three
components increase steadily with temperature, but $\alpha_{zx}^y$ and $\alpha_{yz}^x$ increase much
faster than $\alpha_{xy}^z$. This is the reason why $\alpha_{xy}^z$ is so small even above 300 K.

For the conversion efficiency of the charge and heat current to the spin current not
only the absolute values of the SHC and SNC are important, but also the 
spin Hall angle (SHA) and the spin nernst angle (SNA).
Due to the lack of experimental values for the electrical conductivity 
and thermopower of InBi film, we have just used the bulk values to
estimate the SHA and SNA. Because of the large SHC ($\sigma_{xy}^z$) 
and small charge conductivity~\cite{cooper1964electrical}, the SHA
($\theta_{SH,xy}^{z}$) can reach up to 0.3 (see Table I). Hence, 
the SHA in InBi is close to or even larger than that in the 5$d$
transition metals of Pt, W, and Ta. Similarly, in combination with a large SNC
and low thermalpower, the SNA
can reach up to 0.21 and 0.28 for $\theta_{SN,yz}^{x}$ and $\theta_{SN,zx}^{y}$, 
respectively. It should be noted that the thermalpower in bulk is normally much 
larger than that in film form, so the SNA should also be much larger.

\begin{table}[t]
  \centering
  \caption{SHC, SHA, SNC, and SNA for InBi and Pt. The chemical potentiol is
at charge neutral point, and the temperature is 300 K. The unit for SHC and SNC
are $(\hbar/e) (\Omega\cdot cm)^{-1}$ and $((\hbar/e)(A/m\cdot K))$, respectively.}
    \label{tab:table1}
  \begin{tabular} {rrrrrr}
  \hline
  \hline
                     &                &  SHC   &  SHA   & SNC  & SNA    \\
  \hline
                     &  $X_{xy}^{z}$  &  1100  &  0.3   & 0.2  & 0.04   \\
    InBi             &  $X_{zx}^{y}$  &  300   &  0.09  & 1.2  & 0.28    \\
                     &  $X_{yz}^{x}$  &  200   &  0.06  & 0.9  & 0.21    \\
\hline
    Pt               &  $X_{xy}^{z}$  &  2200  &  0.11  & 3.0  & 0.20     \\
\hline
\hline
  \end{tabular}
\end{table}

At low temperature, the SNC can be understood from the Mott relation as the derivative of the SHC with respect to
energy~\cite{Lee2004,Xiao2006}:
\begin{equation}
\alpha_{ij}^{k}=-\frac{\pi^{2}}{3}\frac{k_{B}^{2}T}{e}\frac{\partial\sigma_{ij}^{k}(E)}{\partial E}
\end{equation}
At a temperature of about 10 K, the SNC from both the Mott relation and SNBC formalism converge to about zero.
From Fig. 3 (c), one can easily see that the ANC from the two different formalisms of formula (2) and (3)
agree very well with each other at low temperatures from 10 to 100 K. As the slopes of the energy-dependent SHC for $\sigma_{zx}^y$
and $\sigma_{yz}^x$ are significantly sharper than that in $\alpha_{xy}^z$, $\alpha_{zx}^y$ and $\alpha_{yz}^x$ are
much larger than $\alpha_{xy}^z$ at temperatures far away from 0 K.

\begin{figure}[htbp]
\begin{center}
\includegraphics[width=0.85\textwidth]{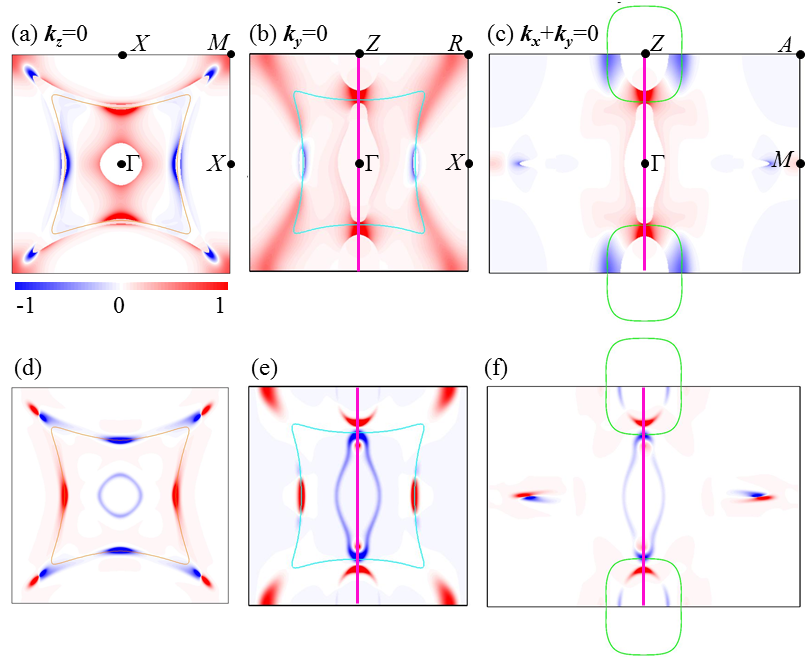}
\end{center}
\caption{
\textbf{Local Berry curvature $\Omega_{xy}^{S,z}(\vec{k})$ and $\Omega_{ xy}^{SN,z}(\vec{k})$
distribution on high symmetry planes}
(a-c) and (d-f) are SBC and SNBC distribution on the three planes of
$k_z$=0, $k_y$=0, and $k_x+k_y$=0 planes, respectively. The nodal
lines are also shown by orange, cyan, green circles, and magenta curves, respectively.
The color bars are in arbitrary units.
}
\label{xyz}
\end{figure}

From Eqs. (3) and (4), one can see that the intrinsic SHC and SNC can be understood as the integral of the SBC and SNBC in the
whole BZ. As the maximum elements are different for the SHC and SNC,
we choose the SBC and SNBC from two different components of $\sigma (\alpha)_{xy}^z$ and $\sigma (\alpha)_{zx}^y$ to observe their
distribution and temperature effect for the evolution from the SBC to SNBC. From the above band structure analysis, we know that the special
band structures mainly focus on the high-symmetry planes of $k_{x,y}$=0, $k_z$=0, and $k_{x}\pm k_{y}$=0, and high-symmetry line of $\Gamma-Z$. Therefore, we have analysed the SBC and SNBC distribution in all the three planes of $k_{x,y}$=0, $k_z$=0, and
$k_{x}+k_{y}$=0.

First, we analysed the component of $\Omega_{xy}^{S,z}$, which contributes to the
large tensor element for the SHC.
Fixing the energy at the charge neutral point, as shown in Fig. 4(a), the large SBC in the $k_z$=0 plane
is mainly dominated by the $\Gamma$-centred nodal ring. In addition, there
exist other hotspots on the $\Gamma$-$M$
line near the $M$ point, which is not exactly located on the nodal ring. This is because the SBC is indeed determined by
the band structure with the SOC. The gapless linear band crossing transforms to a band anti-crossing by the SOC. Comparing the linear
crossing point $K1$ (highlighted by orange circle) in Fig. 2(b) and the corresponding band anti-crossing point $K1'$
in Fig. 2(c), we can find that the $K1'$ is much closer to the $M$ point than $K1$. Therefore, the
hotspots along $\Gamma$-$M$ do not exactly lie on the corner of the gapless nodal ring. In addition, the Fermi level does not
lie in the band gap of the band anti-crossing but slightly cuts the conduction band, which also implies the strong entanglement
between the conduction and valence band around the band anti-crossing points.

Meanwhile, in the $k_y$=0 plane, there are two types of hotspots, one from the nodal ring in the $k_y$=0 plane and the other on the
line $Z$-$\Gamma$-$Z$. The $\Omega_{xy}^{S,z}$ from the nodal line on $Z$-$\Gamma$-$Z$ is much larger, which is related to
the magenta bands in Figs. 2(c--d). There are also two types of hotspots in the $k_x+k_y$=0 plane,
one from $Z$-$\Gamma$-$Z$ as that in the $k_y$=0 plane and the other from the $Z$-point-centred nodal ring. From Figs. 4(b--c),
one can find some empty parts with a zero SBC and shape edge transition, such as the $\Gamma$- and $Z$-centred areas.
This is due to the shape evolution of the band occupation, such as the band around the $\Gamma$ point, as shown in Fig. 2(d). For the $k$ point from
$R$ to $\Gamma$, the valence band changes from an occupied band to a non-occupied band after a transition $k$ point (highlighted
by the black dot). Similar behaviour is also exhibited for the other areas with an empty SBC.

After taking the temperature effect into consideration, the SNBC ($\Omega_{xy}^{SN,z}$) also mainly
originates from the four classes of nodal lines as that for the SBC.
The main difference is that the shape edge transitions in the empty SBC
areas are replaced by the hotspot of the SNBC (see Figs. 4(d--f)).
This is because the temperature effect smoothens considerably the transition of
the band occupation, and the shape transition of the
SBC around the edge of the empty area yields a large SNBC.

\begin{figure}[htbp]
\begin{center}
\includegraphics[width=0.85\textwidth]{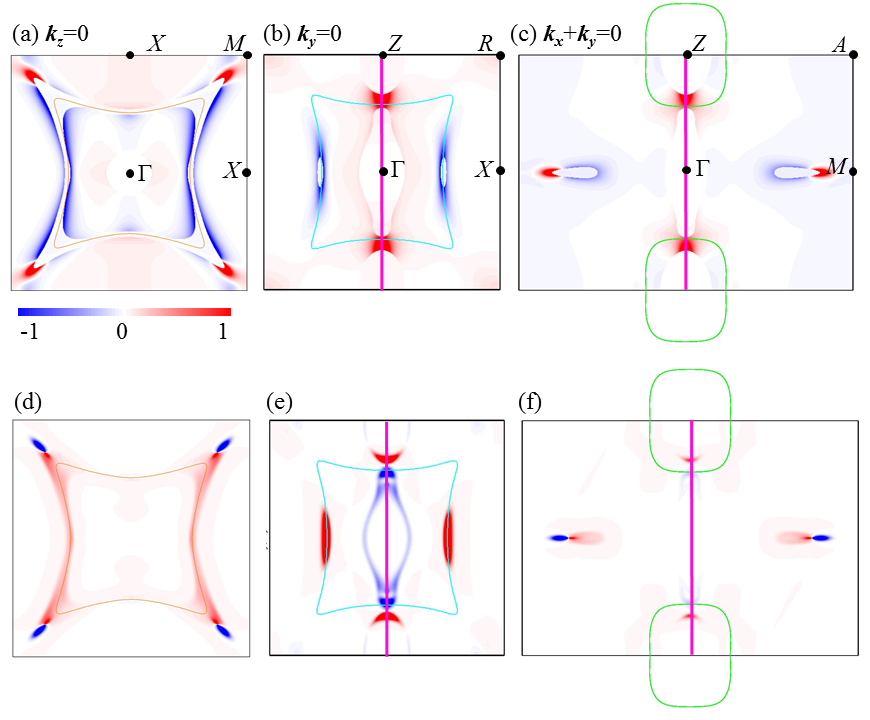}
\end{center}
\caption{
\textbf{Local Berry curvature distribution on high symmetry planes for the
component of $\Omega_{zx}^{S,y}(\vec{k})$ and $\Omega_{zx}^{SN,y}(\vec{k})$}
(a-c) and (d-f) are SBC and SNBC distribution on the three planes of
$k_z$=0, $k_y$=0, and $k_x+k_y$=0 planes, respectively. The nodal
lines are also shown by orange, cyan, green circles, and magenta curves, respectively.
The color bars are in arbitrary units.
}
\label{zxy}
\end{figure}

Similar to $\Omega_{xy}^{S,z}$, the component of $\Omega_{zx}^{S,y}$ is
 dominated by the four classes of nodal lines, but the volume
of the negative part is larger than that of $\Omega_{xy}^{S,z}$ (see Figs. 5(a--c)),
leading to a relatively smaller absolute value for the SHC tensor element $\sigma_{zx}^y$.
As the negative SNBC($\Omega_{zx}^{NS,y}$) primarily originates from
the sharp transition of the SBC around the edge of the zone with an empty SBC, the
small magnitude of $\Omega_{zx}^{S,y}$ generates a relatively small SNBC
for the same tensor component. Comparing Figs. 4 and 5,
the $\Gamma$-centred negative hot rings for $\Omega_{xy}^{SN,z}$ in Figs. 4(d)
and (f) are almost invisible in Figs. 5(d) and (f) for $\Omega_{zx}^{SN,y}$.
Moreover, the $\Gamma$-centred negative hot ring for $\Omega_{zx}^{SN,y}$
in Fig. 5(e) is also considerably weaker than $\Omega_{xy}^{SN,z}$ in Fig. 5(d).
Because both components of $\alpha_{xy}^z$ and $\alpha_{zx}^y$ are positive
values, the smaller negative $\Omega_{zx}^{NS,y}$ directly leads to a larger $\alpha_{zx}^y$
in comparison to $\alpha_{xy}^z$.

In summary, from our calculations we predict a large SHC and SNC in the $p$-band semimetal InBi.
Due to the contribution of the nodal lines in the band structure of InBi, the $\sigma_{xy}^z$ component of the SHC can reach up to about $1100$
($\left(\hbar/e\right)\left(\Omega\cdot cm\right)^{-1}$), the only SHC above 1000
($\left(\hbar/e\right)\left(\Omega\cdot cm\right)^{-1}$) in the reported $p$-band systems.
In contrast to the intuition that a large SHC is always accompanied by a large SNC, we find that
the largest value for the SHC and SNC appear in a different third-order tensor element. The size of the SNC is mainly dependent on the breaking of the balance of the SHC distribution in energy space and gives the
largest SNC of about $1.2$ $((\hbar/e)(A/m\cdot K))$ at the tensor element $\alpha_{zx}^y$, which is close to that in the 5$d$
transition metal platinum. 
These results on InBi provide more general guiding principles to obtain a strong SHE and SNE in commercially more attractive $p$-band compounds and establishes a more comprehensive understanding of the relationship between the SHE and SNE in these materials.

\textbf{Method}
First-principles calculations were performed using the localized atomic
orbital basis and the full potential, as implemented in the Full Potential Local Orbital code (FPLO)~\cite{Koepernik1999}.
Exchange and correlations are considered in the generalized gradient approximation,
following the Perdew--Burke--Ernzerhof parametrization scheme~\cite{perdew1996}.
We employed the experimental lattice constants in all our calculations~\cite{Sandy2017}.
We constructed a high-symmetry tight binding Hamiltonian by projecting the Bloch states onto
atomic orbital-like Wannier functions and computed the SHC by the linear-response
Kubo formula approach~\cite{Xiao2010, Sinova2015}:
\begin{equation}
\begin{aligned}
\sigma_{ij}^{k}={e} \int_{_{BZ}}\frac{d\vec{k}}{(2\pi)^{3}}\underset{n}{\sum}f_{n\vec{k}}\Omega_{n,ij}^{S,k}(\vec{k}), \\
\Omega_{n,ij}^{S,k}(\vec{k})=-2Im\underset{n'\neq n}{\sum}\frac{\langle n\vec{k}| J_{i}^{k}|n'\vec{k} \rangle \langle n'\vec{k}| v_{j}|n\vec{k}\rangle}{(E_{n\vec{k}}-E_{n'\vec{k}})^{2}}
\end{aligned}
\label{SHC}
\end{equation}
where $f_{n\vec{k}}$ is the Fermi--Dirac distribution for the $n$-th band.
$J_{i}^{k}=\frac{1}{2}\left\{ \begin{array}{cc}
{v_{i}}, & {s_{k}}\end{array}\right\} $
is the spin current operator with spin operator ${s}$, velocity operator
${v_{i}}$, and $i,j,k=x,y,z$.
$| n\vec{k} \rangle$ is the eigenvector for the Hamiltonian
${H}$ at eigenvalue $E_{n\vec{k}}$.
$\Omega_{n, ij}^{S,k}(\vec{k})$
is referred to as the SBC for the $n$-th band at point $\vec{k}$
in analogy to the ordinary Berry curvature. The SNC is calculated via
\begin{equation}
\begin{aligned}
  \alpha_{ij}^{k}=\frac{1}{T} \int_{BZ} \frac{d^3k}{(2\pi)^3} \sum_n \Omega_{n,ij}^{S,k}(\vec{k})[(E_{n}-E_F)f_{n \vec{k}}+k_BT\ln{(1+\exp{\frac{E_{n}-E_F}{-k_BT}})}],
\end{aligned}
\end{equation}
A $500 \times 500 \times 500$ $k$-grid in the BZ was used for the integral of the SHC and SNC.
For convenience, we call $\Omega_{n,ij}^{SN,k}$ =$\Omega_{n,ij}^{S,k}(\vec{k})[(E_{n}-E_F)f_{n \vec{k}}+k_BT\ln{(1+\exp{\frac{E_{n}-E_F}{-k_BT}})}]$
the SNBC.

The BHZ model~\cite{Bernevig_2006} around the $\Gamma$ point is written in the form of
\begin{equation}
\begin{aligned}
H_{eff}(\vec{k})=\left(\begin{array}{cc}
H(\vec{k})\\
 & H^{*}(-\vec{k})
\end{array}\right) \\
H(\vec{k})=\varepsilon(\vec{k})+d_{i}(\vec{k})\sigma_{i}
\end{aligned}
\end{equation}
where $\sigma_i$ are the Pauli matrices, $d_1+id_2=A(k_x+ik_y)$, and $d_3=M-B(k_{x}^{2}+k_{y}^{2})$.
In our calculations, for Figs. 1(a--c), we have set the parameters of
$A=-0.1$ eV ${\AA}$, $B=-0.5$ eV ${\AA}^{2}$, and $M=0.1$ eV.

In order to obtain a non-zero SNC, we break the balance of the SHC by reducing the symmetry:
\begin{equation}
\begin{aligned}
H_{eff}(\vec{k})=\left(\begin{array}{cc}
H^{'}(\vec{k})\\
 & H^{*'}(-\vec{k})
\end{array}\right) \\
H'(\vec{k})=d_{1}^{'}\sigma_{1}+d_{2}^{'}\sigma_{2}+\left(\begin{array}{cc}
M_{1}-(B_{1}k_{x}^{2}+B_{2}k_{y}^{2}) & 0\\
0 & -[M_{2}-(B_{2}k_{x}^{2}+B_{1}k_{y}^{2})]
\end{array}\right)
\end{aligned}
\end{equation}
where $d_{1}^{'}+id_{2}^{'}=A_{1}k_x+iA_{2}k_y$. The parameters are $A_1$=0.05 eV ${\AA}$,
$A_2$= 0.1 eV ${\AA}$, $B_1$=-1 eV ${\AA}^{2}$, $B_2$=-0.5 eV ${\AA}^{2}$, $M_1$=-0.3 eV,
and $M_1$=-0.5 eV.

To perform the integral in the whole BZ, we projected the continuous $\vec{k}\cdot\vec{p}$ model
to the lattice by the replacement of $k_{i}=(1/a)sin(ak_i)$ and $k_{i}^2=(2/a^{2})(1-cos(ak_i))$ with
lattice constant $a$=1 $\AA$

\begin{acknowledgments}
This work was financially supported by the ERC
Advanced Grant No.
291472 ``Idea Heusler'' and ERC
Advanced Grant No. 742068 ``TOPMAT''. Y.Z., J.vdB. and C.F. thank financial support by the
German Research Foundation (DFG, SFB 1143 A05).
\end{acknowledgments}

\bibliographystyle{ieeetr}
\bibliography{shc_ref}

\end{document}